\documentclass[12pt]{iopart}
\usepackage{graphicx}

\begin{document}

\title[Resonance Production]{Resonance Production}

\author{Patricia Fachini
\footnote[3]{To whom correspondence should be addressed
(pfachini@bnl.gov)} }

\address{Brookhaven National Laboratory,
Upton, NY, 11973, USA}

\begin{abstract}
Recent results on $\rho(770)^0$, $K(892)^{*0}$, $f_{0}(980)$,
$\phi(1020)$, $\Delta(1232)^{++}$, and $\Lambda(1520)$ production
in A+A and $p+p$ collisions at SPS and RHIC energies are
presented. These resonances are measured via their hadronic decay
channels and used as a sensitive tool to examine the collision
dynamics in the hadronic medium through their decay and
regeneration. The modification of resonance mass, width, and shape
due to phase space and dynamical effects are discussed.
\end{abstract}




\section{Introduction}
The in-medium modification of vector mesons has been proposed as a
possible signal of a phase transition of nuclear matter to a
deconfined plasma of quarks and gluons in relativistic heavy-ion
collisions \cite{1}.

Even in the absence of a phase transition, at normal nuclear
density, the modification of vector meson properties are expected
to be measurable. Effects such as phase space
\cite{2,3,4,5,6,7,8,9,10} and dynamical interactions with matter
\cite{4,6,8,11} may modify the resonance mass, width, and shape.
These modifications of the vector meson properties take place
close to kinetic freeze-out, in a dilute hadronic gas at late
stages of heavy-ion collisions. At such low matter density, the
proposed modifications are expected to be small, but observable.
The effects of phase space due to the rescattering of hadrons and
Bose-Einstein correlations between the daughters from the
resonance decay and the hadrons in the surrounding matter are
present in $p+p$ \cite{3,4,6,10} and Au+Au \cite{2,4,5,6,7,8,9,11}
collisions. The interference between different pion scattering
channels can effectively distort the line shape of resonances
\cite{12}. Dynamical effects due to vector mesons interacting with
the surrounding matter are also expected to be present in both
systems, and have been evaluated for the latter \cite{4,6,8,11}.

Those resonances that decay before kinetic freeze-out may not be
reconstructed due to the rescattering of the daughter particles.
In this case, the resonance survival probability is relevant and
depends on the time between chemical and kinetic freeze-outs, the
source size, and the $p_T$ of the resonance. On the other hand,
after chemical freeze-out, elastic interactions may increase the
resonance population compensating for the ones that decay before
kinetic freeze-out. This resonance regeneration depends on the
hadronic cross-section of their daughters. For example, the $K^*$
regeneration depends on the $K\pi$ cross section ($\sigma$) while
the rescattering of the daughter particles depends on
$\sigma_{\pi\pi}$ and $\sigma_{\pi p}$, which are considerably
larger (factor $\sim$5) than $\sigma_{K\pi}$ \cite{13,14,15}. In
the case of the $\Delta^{++}$, the regeneration probability is
expected to be higher than the rescattering of the daughters since
$\sigma_{\pi p} > \sigma_{\pi\pi}$ \cite{13,15}. Thus, the study
of resonances can provide an independent probe of the time
evolution of the source from chemical to kinetic freeze-outs and
yield detailed information on hadronic interaction at later
stages.

\section{Results}

The $\rho(770)^0$ \cite{16,17}, $K(892)^{*0}$ \cite{17},
$f_{0}(980)$ \cite{18}, $\phi(1020)$ \cite{19},
$\Delta(1232)^{++}$ \cite{17}, $\Lambda(1520)$ \cite{20}, and
$\Sigma(1385)$ \cite{20,21} production were measured via their
hadronic decay channels at midrapidity ($|y| \!\leq\! 0.5$) in
Au+Au and $p+p$ collisions at $\sqrt{s_{_{NN}}}\!=\!$ 200 GeV
using the STAR detector at RHIC. The $\phi$ \cite{22} meson was
also studied via its hadronic decay channel at $|y| \!\leq\! 0.5$
in Au+Au collisions using the PHENIX detector at RHIC. At SPS, the
$\phi$ \cite{23} meson was recently measured via its hadronic
decay channel by NA49 at $0.0 \!\leq\! y \!\leq\! 2.0$ for the top
7$\%$ of the inelastic hadronic Pb+Pb cross-section at
$E_{beam}\!=\!$ 80, 40, 30, and 20 $A$GeV ($\sqrt{s_{_{NN}}}\!=\!$
12.32, 8.76, 7.62, and 6.27 GeV, respectively).

\begin{figure}[htb]
\begin{minipage}[t]{80mm}
\includegraphics[height=14pc,width=18pc]{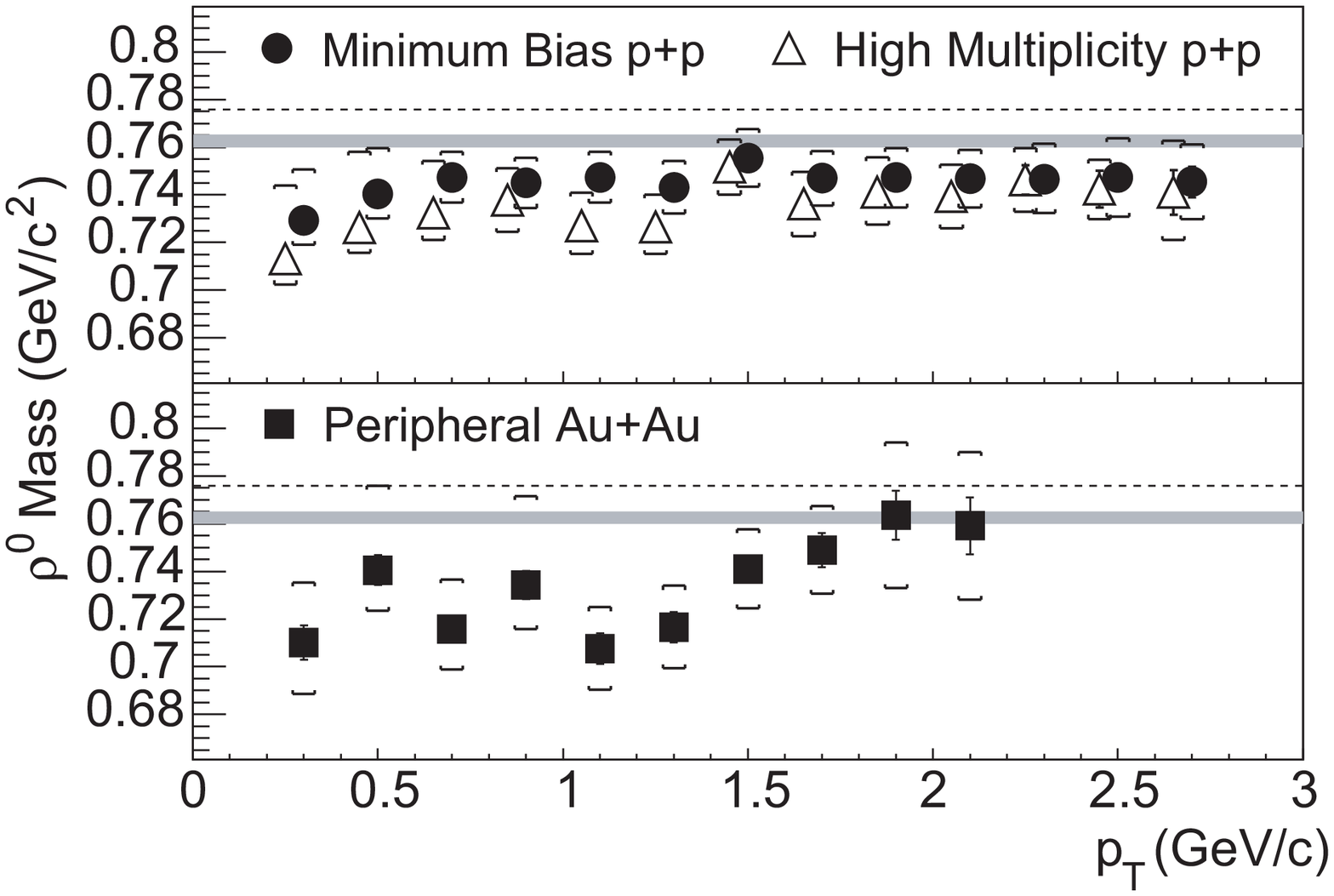}
\end{minipage}
\hspace{\fill}
\begin{minipage}[t]{85mm}
\includegraphics[height=14pc,width=18pc]{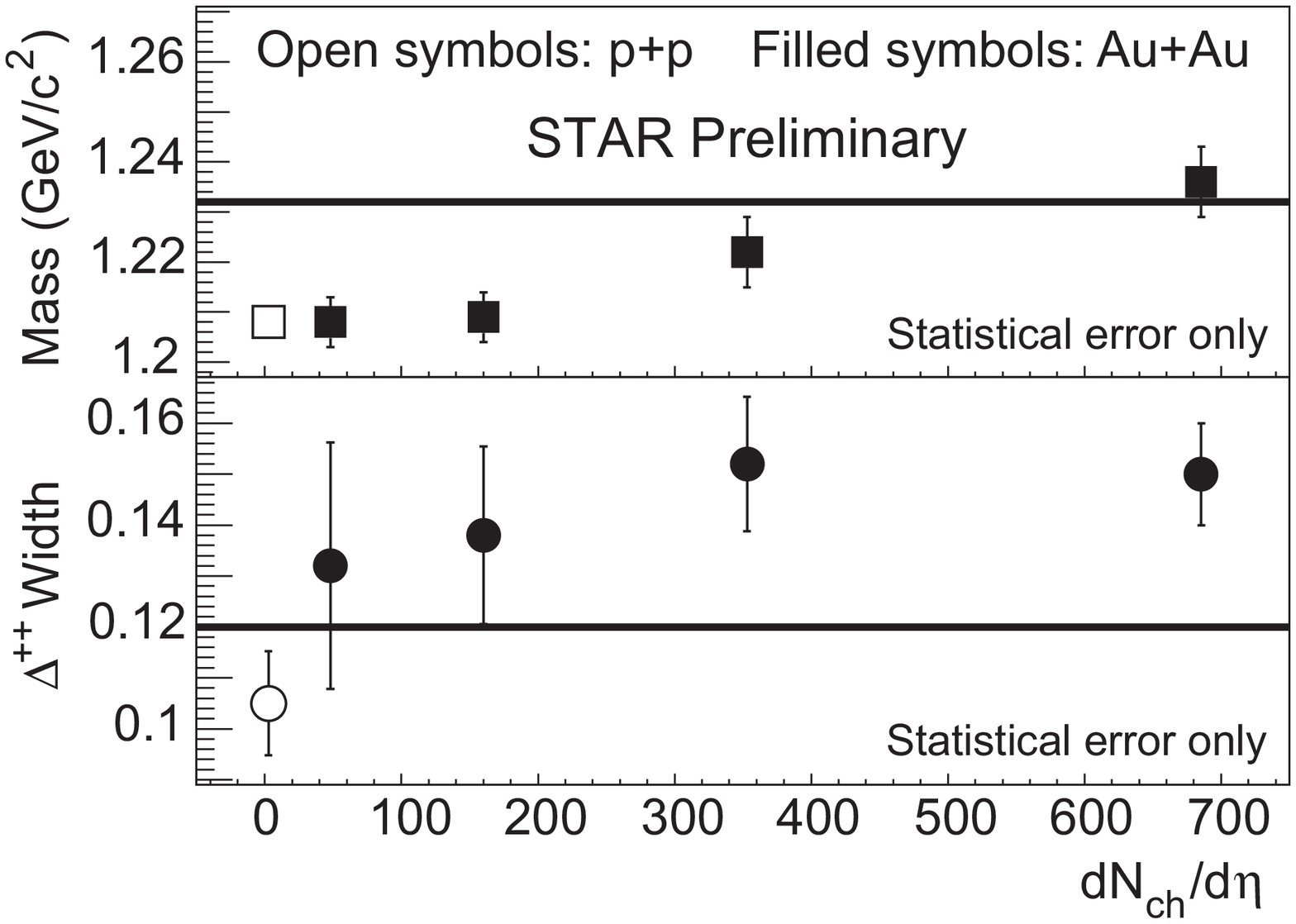}
\end{minipage}
\caption{Left panel: The $\rho^0$ mass as a function of $p_T$. The
error bars indicate the systematic uncertainty. The dashed lines
represent the average of the $\rho^0$ mass measured in $e^+e^-$
\cite{15}. The shaded areas indicate the $\rho^0$ mass measured in
$p+p$ collisions \cite{24}. The open triangles have been shifted
downward on the abscissa for clarity. Right panel: $\Delta^{++}$
mass (top) and width (bottom) as a function of $dN_{ch}/d\eta$.
The solid lines correspond to the average of the $\Delta^{++}$
mass and width reported in \cite{15}. The errors shown are
statistical only.} \label{fig:Mass1}
\end{figure}

\begin{figure}[htb]
\begin{minipage}[t]{80mm}
\includegraphics[height=14pc,width=18pc]{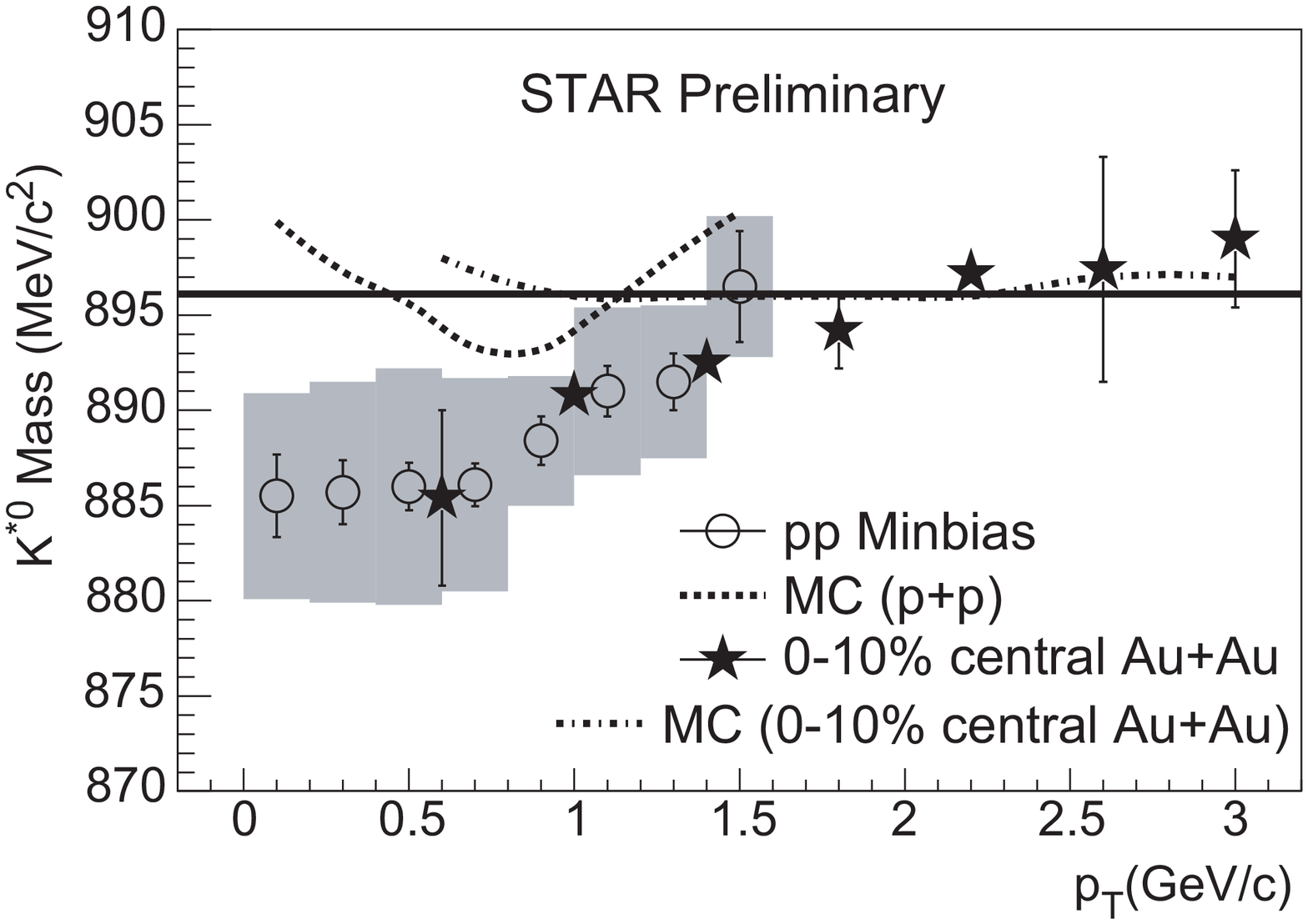}
\end{minipage}
\hspace{\fill}
\begin{minipage}[t]{85mm}
\includegraphics[height=14pc,width=18pc]{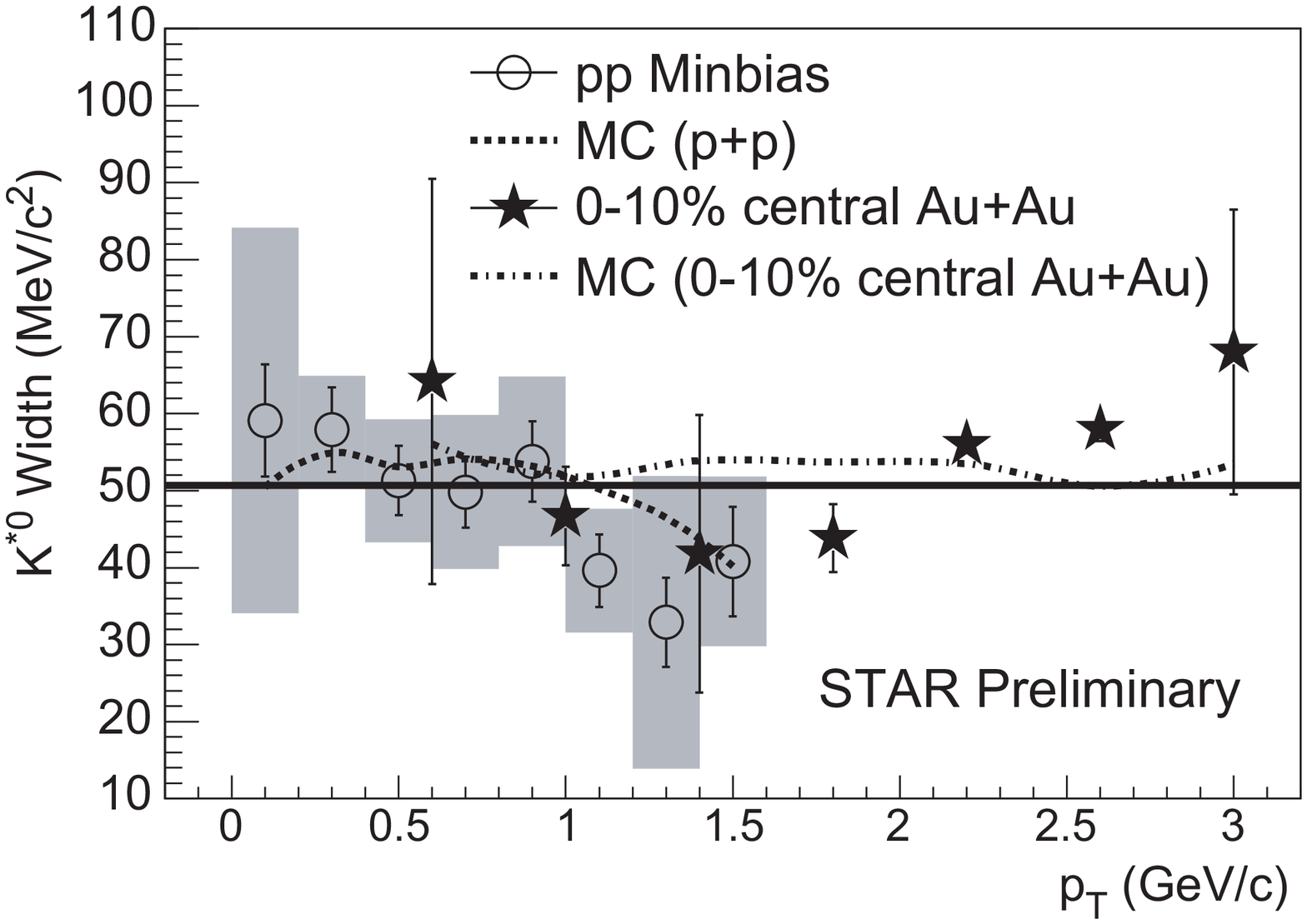}
\end{minipage}
\caption{The $K^{*0}$ mass (left panel) and width (right panel) as
a function of $p_T$. The grey shaded boxes are the systematic
uncertainties in minimum bias $p+p$. The solid lines correspond to
the average of the $K^{*0}$ mass and width reported in \cite{15}.
The dotted and dashed lines are the results from the Monte Carlo
simulations, which accounts for detector effects and kinematic
cuts, for the Au+Au and $p+p$ measurements, respectively. The
errors shown for the Au+Au measurement are statistical only.}
\label{fig:Mass2}
\end{figure}

The $\rho^0$ mass is shown as a function of $p_T$ in
Fig.~\ref{fig:Mass1} for peripheral Au+Au (40-80$\%$ of the
inelastic hadronic cross-section), high multiplicity $p+p$ (top
10$\%$ of the minimum bias $p+p$ multiplicity distribution for
$|\eta| \!<\!$ 0.5), and minimum bias $p+p$ interactions. The
$\rho^0$ mass was obtained by fitting the data to a $p$-wave
Breit-Wigner function times the phase space (BW$\times$PS)
described in \cite{16}. Figure~\ref{fig:Mass1} also depicts the
$\Delta^{++}$ mass and width as a function of charged particle
multiplicity ($dN_{ch}/d\eta$) for minimum bias $p+p$ and Au+Au
collisions. The $\Delta^{++}$ mass was obtained by fitting the
data to the BW$\times$PS function \cite{16} and a gaussian
function representing the residual background described in
\cite{17}. Figure~\ref{fig:Mass2} also shows the $K^{*0}$ mass and
width as a function of $p_T$ for central Au+Au (top 10$\%$ of the
hadronic cross-section) and minimum bias $p+p$ interactions. The
$K^{*0}$ mass was obtained by fitting the data to the BW$\times$PS
function \cite{16} and a linear function representing the residual
background described in \cite{17}.

The $\Delta^{++}$ mass measured in minimum bias $p+p$ and Au+Au
collisions for low values of $dN_{ch}/d\eta$ is lower than the
value reported in \cite{15} within statistical errors even after
correcting for phase space (PS). Both mass and width increase as a
function of centrality, and while the $\Delta^{++}$ mass is in
agreement with the value reported in \cite{15} for large values of
$dN_{ch}/d\eta$ within statistical errors, its width is larger
than the average reported in \cite{15}. The increase of the
$\Delta^{++}$ mass as a function of centrality was predicted in a
recent calculation that introduces the in-medium modification of
the $\Delta^{++}$ \cite{4}. Both $\rho^0$ and $K^{*0}$ masses
increase as a function of $p_T$ and are systematically lower than
the value reported in \cite{15}. The $K^{*0}$ width is in
agreement with the average presented in \cite{15}. The $\rho^0$
mass measured in peripheral Au+Au collisions is lower than the
minimum bias $p+p$ measurement. The mass for high multiplicity
$p+p$ interactions is lower than for minimum bias $p+p$
interactions for all $p_T$ bins, indicating that the $\rho^0$ mass
is also multiplicity dependent. Recent calculations are not able
to reproduce the $\rho^0$ mass measured in peripheral Au+Au
collisions without introducing in-medium modification of the
$\rho^0$ meson \cite{4,5,6,7,8,9,11}.

Previous observations of the $\rho$ meson in $e^+e^-$
\cite{24,25,26} and $p+p$ interactions \cite{27} indicate that the
$\rho^0$ line shape is considerably distorted from a $p$-wave
Breit-Wigner function. A mass shift of $-$30 MeV/$c^{2}$ or larger
was observed in $e^+e^-$ collisions at $\sqrt{s}$ $\!=\!$ 90 GeV
\cite{24,25,26}. In the $p+p$ measurement at $\sqrt{s}$ $\!=\!$
27.5 GeV \cite{27}, a $\rho^0$ mass of 0.7626 $\!\pm\!$ 0.0026
GeV/$c^2$ was obtained from a fit to a relativistic $p$-wave
Breit-Wigner function times the phase space \cite{27}. This result
is the only $p+p$ measurement used in the hadro-produced $\rho^0$
mass average reported in \cite{15}.

The $\rho^0$ \cite{16,17} and $f_0$ \cite{18} measurements do not
have sufficient sensitivity to permit a systematic study of the
their widths and the mass of the later. The $\phi$ \cite{19,22,23}
and the $\Lambda(1520)$ \cite{17} masses and widths are in
agreement with the values reported in \cite{15}.

\begin{figure}[htb]
\begin{minipage}[t]{80mm}
\includegraphics[height=14pc,width=18pc]{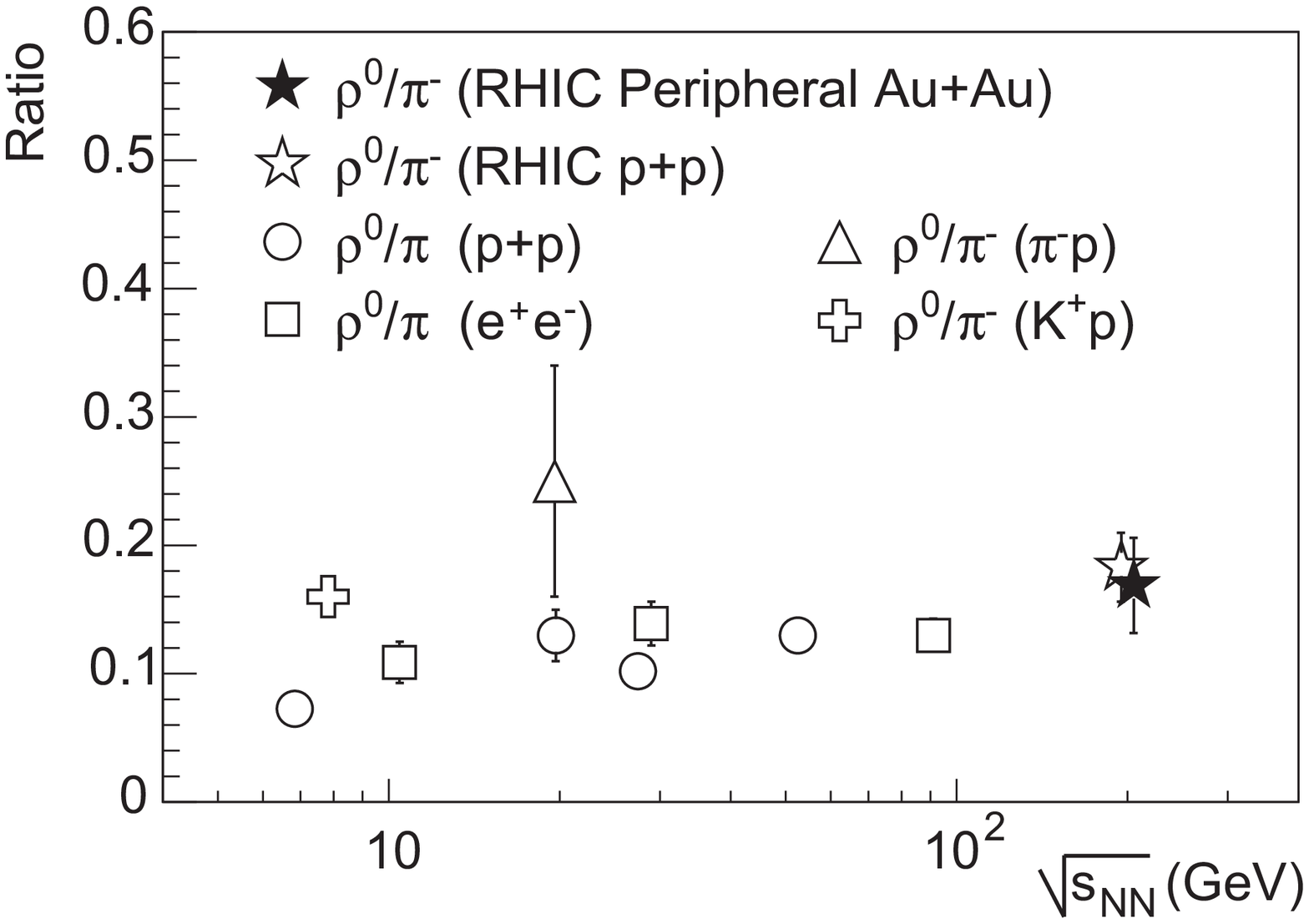}
\end{minipage}
\hspace{\fill}
\begin{minipage}[t]{85mm}
\includegraphics[height=13.4pc,width=18pc]{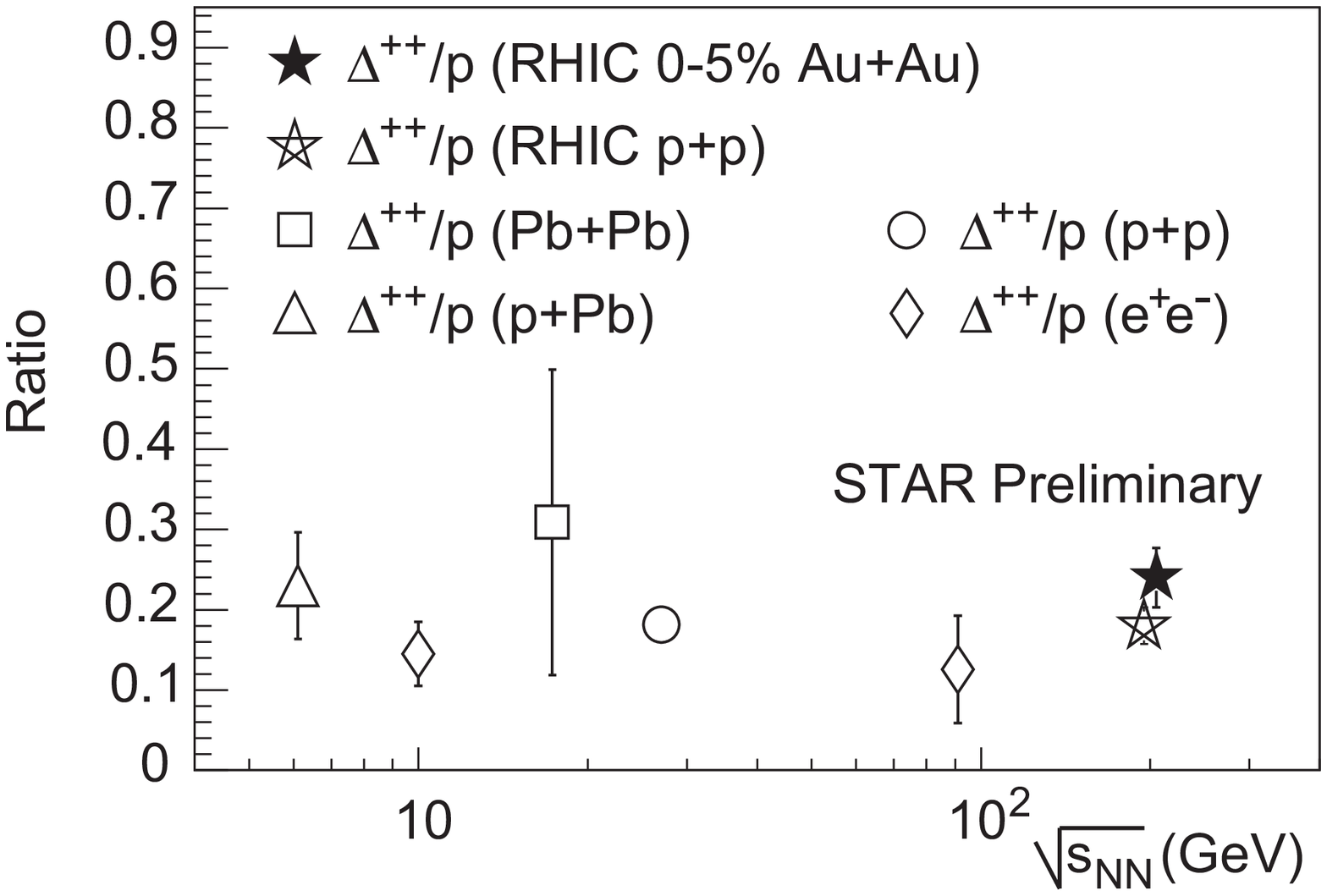}
\end{minipage}
\caption{Left: $\rho^0/\pi$ ratios in $e^+e^-$ \cite{28,29,30},
$p+p$ \cite{27,31,32,33}, $K^+p$ \cite{34}, and $\pi^-p$ \cite{35}
interactions at different c.m. system energies compared to the
recent STAR measurement at $|y| \!<\!$ 0.5 for minimum bias $p+p$
and peripheral Au+Au collisions at $\sqrt{s_{_{NN}}}\!=\!$ 200 GeV
\cite{16,17}. Right: $\Delta^{++}/p$ ratios in $e^+e^-$ \cite{36},
$p+p$ \cite{27}, $p$+Pb \cite{37}, and Pb+Pb \cite{38}
interactions at different c.m. system energies compared to the
recent STAR measurement at $|y| \!<\!$ 0.5 for minimum bias $p+p$
and for the top 5$\%$ of the inelastic hadronic Au+Au
cross-section at $\sqrt{s_{_{NN}}}\!=\!$ 200 GeV \cite{17}. The
errors on the ratios at $\sqrt{s_{_{NN}}} \!=\!$ 200 GeV are the
quadratic sum of the statistical and systematic errors. The ratios
at $\sqrt{s_{_{NN}}} \!=\!$ 200 GeV are offset from one another
for clarity.} \label{fig:Ratio1}
\end{figure}

\begin{figure}[htb]
\begin{minipage}[t]{80mm}
\includegraphics[height=14pc,width=18pc]{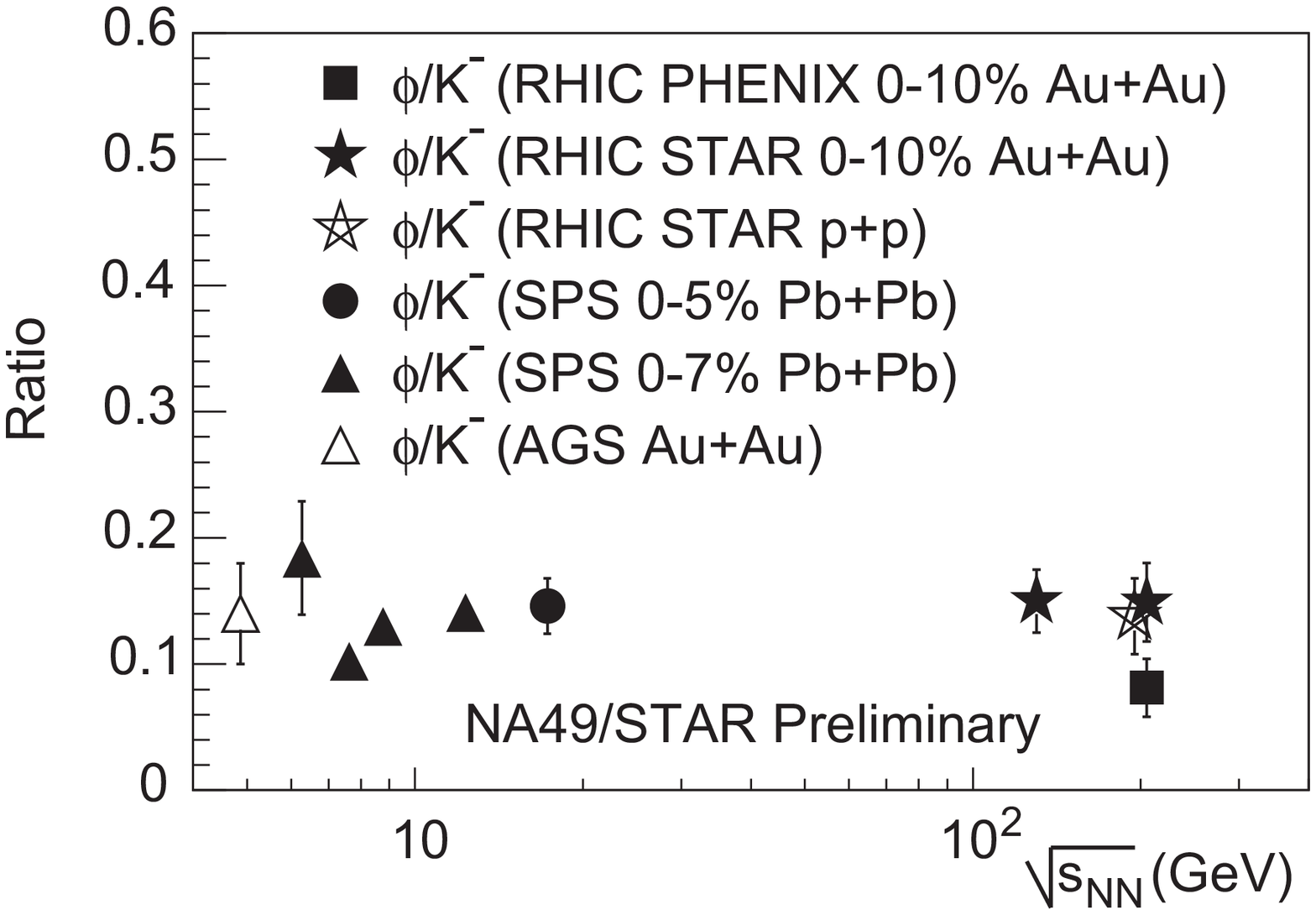}
\end{minipage}
\hspace{\fill}
\begin{minipage}[t]{85mm}
\includegraphics[height=14pc,width=18pc]{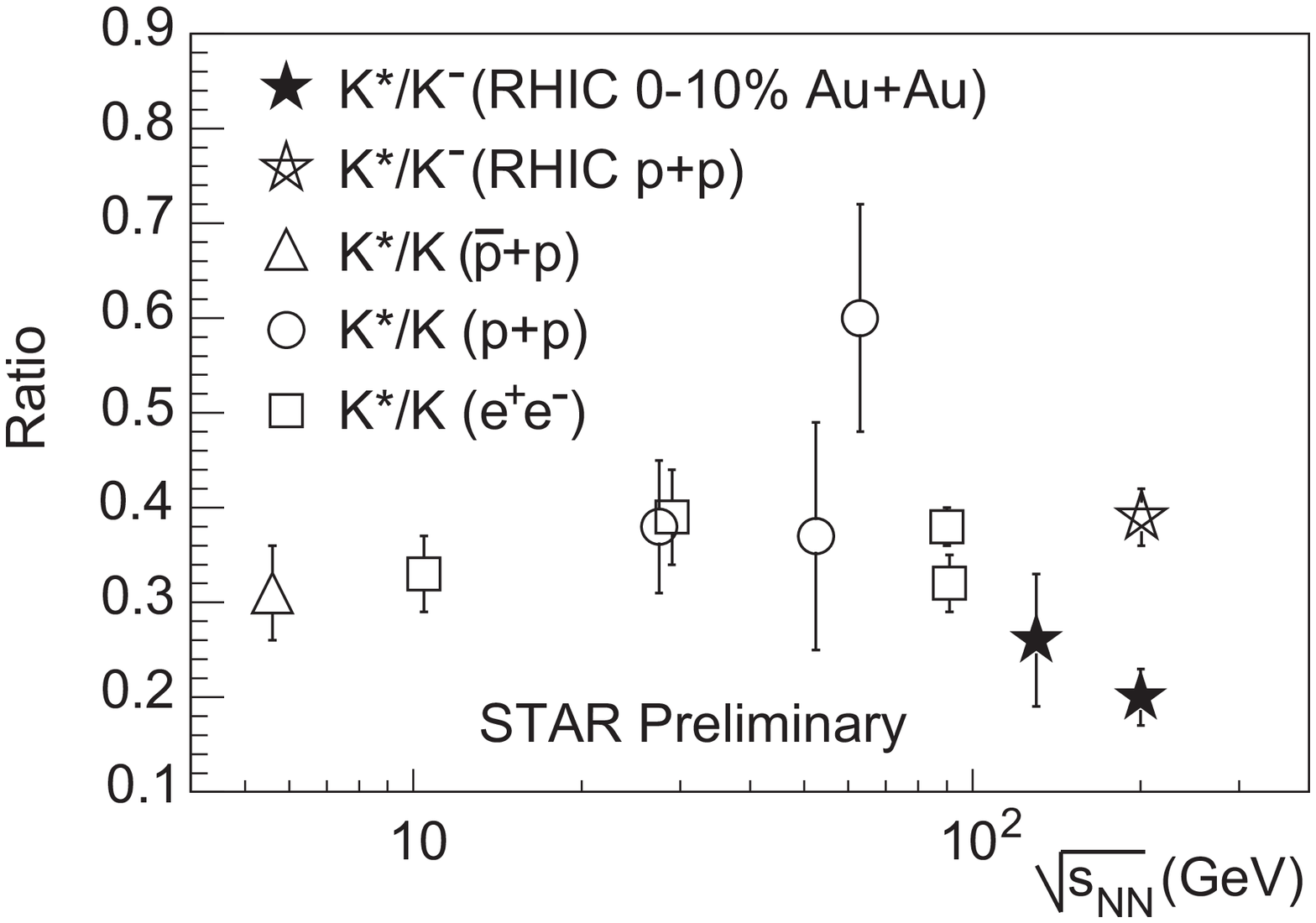}
\end{minipage}
\caption{Left: $\phi/K^-$ ratios in Au+Au collisions at 4.87 GeV
\cite{39} and in Pb+Pb collisions for the top 5$\%$ of the
inelastic hadronic cross-section at 17.27 GeV \cite{40} compared
to the recent NA49 results for the top 7$\%$ of the inelastic
hadronic Pb+Pb cross-section at $\sqrt{s_{_{NN}}}\!=\!$ 12.32,
8.76, 7.62, and 6.27 GeV \cite{23}, the recent STAR measurements
at $|y| \!<\!$ 0.5 for minimum bias $p+p$ and for the top 10$\%$
of the inelastic hadronic Au+Au cross-section at
$\sqrt{s_{_{NN}}}\!=\!$ 200 GeV \cite{19}, and the recent PHENIX
measurement at $|y| \!<\!$ 0.5 for the top 10$\%$ of the inelastic
hadronic Au+Au cross-section at $\sqrt{s_{_{NN}}}\!=\!$ 200 GeV
\cite{22}. The $\phi/K^-$ ratios at $\sqrt{s_{_{NN}}} \!=\!$ 200
GeV are offset from one another for clarity. Right: $K^{*0}/K$
ratios in $e^+e^-$ \cite{28,29,30,41}, $p+p$ \cite{27,33,42}, and
$\bar{p}+p$ \cite{43} interactions at different c.m. system
energies compared to the recent STAR measurement at $|y| \!<\!$
0.5 for minimum bias $p+p$ and for the top 10$\%$ of the inelastic
hadronic Au+Au cross-section at $\sqrt{s_{_{NN}}}\!=\!$ 200 GeV
\cite{17}. The STAR measurements of the $\phi/K^-$ and
$K^{*0}/K^-$ ratios at $|y| \!<\!$ 0.5 for the top 10$\%$ of the
inelastic hadronic Au+Au cross-section at $\sqrt{s_{_{NN}}}\!=\!$
130 GeV \cite{44,45} are also depicted. The errors on the ratios
at $\sqrt{s_{_{NN}}} \!=\!$ 200 GeV are the quadratic sum of the
statistical and systematic errors.} \label{fig:Ratio2}
\end{figure}

The $\rho^0/\pi$ ratios as a function of c.m. system energy are
depicted in Fig.~\ref{fig:Ratio1}. The $\rho^0/\pi$ ratios are
from measurements in $e^+e^-$ \cite{28,29,30}, $p+p$
\cite{27,31,32,33}, $K^+p$ \cite{34}, and $\pi^-p$ \cite{35}
interactions at different c.m. system energies compared to the
recent STAR measurement at $|y| \!<\!$ 0.5 for minimum bias $p+p$
and peripheral Au+Au collisions \cite{16,17} at
$\sqrt{s_{_{NN}}}\!=\!$ 200 GeV. The $\Delta^{++}/p$ ratios from
measurements in $e^+e^-$ \cite{36}, $p+p$ \cite{27}, $p$+Pb
\cite{37}, and Pb+Pb \cite{38} interactions are also shown in
Fig.~\ref{fig:Ratio1} and compared to the recent STAR measurement
at $|y| \!<\!$ 0.5 for minimum bias $p+p$ and for the top 5$\%$ of
the inelastic hadronic Au+Au cross-section \cite{17} at
$\sqrt{s_{_{NN}}}\!=\!$ 200 GeV. Figure~\ref{fig:Ratio2} depicts
the $\phi/K^-$ ratios as a function of $\sqrt{s_{_{NN}}}$. The
ratios are from measurements in Au+Au collisions at 4.87 GeV
\cite{39} and in Pb+Pb collisions for the top 5$\%$ of the
inelastic hadronic cross-section at 17.27 GeV \cite{40}, which are
compared to the recent NA49 results for the top 7$\%$ of the
inelastic hadronic Pb+Pb cross-section at $\sqrt{s_{_{NN}}}\!=\!$
12.32, 8.76, 7.62, and 6.27 GeV \cite{23}, the recent STAR
measurements at $|y| \!<\!$ 0.5 for minimum bias $p+p$ and for the
top 10$\%$ of the inelastic hadronic Au+Au cross-section at
$\sqrt{s_{_{NN}}}\!=\!$ 200 GeV \cite{19}, and the recent PHENIX
measurement at $|y| \!<\!$ 0.5 for the top 10$\%$ of the inelastic
hadronic Au+Au cross-section at $\sqrt{s_{_{NN}}}\!=\!$ 200 GeV
\cite{22}. For the top 10$\%$ of the inelastic hadronic Au+Au
cross-section at $\sqrt{s_{_{NN}}}\!=\!$ 200 GeV, while the $\phi$
inverse slope obtained from STAR \cite{19} and PHENIX \cite{22}
are in agreement, there is a factor of 2 difference in the
normalization that corresponds to a 1.25$\sigma$ effect. The
difference in the $\phi$ measurement between STAR and PHENIX is
under investigation. The $K^{*0}/K$ ratios from measurements in
$e^+e^-$ \cite{28,29,30,41}, $p+p$ \cite{27,33,42}, and
$\bar{p}+p$ \cite{43} interactions at different c.m. system
energies are also shown in Fig.~\ref{fig:Ratio2} and compared to
the recent STAR measurement at $|y| \!<\!$ 0.5 for minimum bias
$p+p$ and for the top 10$\%$ of the inelastic hadronic Au+Au
cross-section at $\sqrt{s_{_{NN}}}\!=\!$ 200 GeV \cite{17}. The
STAR measurements of the $\phi/K^-$ and $K^{*0}/K^-$ ratios at
$|y| \!<\!$ 0.5 for the top 10$\%$ of the inelastic hadronic Au+Au
cross-section at $\sqrt{s_{_{NN}}}\!=\!$ 130 GeV \cite{44,45} are
also depicted in Fig.~\ref{fig:Ratio2}.

The $\rho^0/\pi$, $K^{*0}/K$, $\phi/K^-$, and $\Delta^{++}/p$
ratios shown in Fig.~\ref{fig:Ratio1} and Fig.~\ref{fig:Ratio2} do
not present a strong dependence on the colliding system or the
c.m. system energy, with the exception of the $K^{*0}/K^-$ ratio
at $\sqrt{s_{_{NN}}}\!=\!$ 200 GeV. In this case, the $K^{*0}/K^-$
ratio for the top 10$\%$ of the inelastic hadronic Au+Au
cross-section is lower than the minimum bias $p+p$ measurement at
the same c.m. system energy by a factor of 2.

\begin{figure}[htb]
\begin{minipage}[t]{80mm}
\includegraphics[height=14pc,width=19pc]{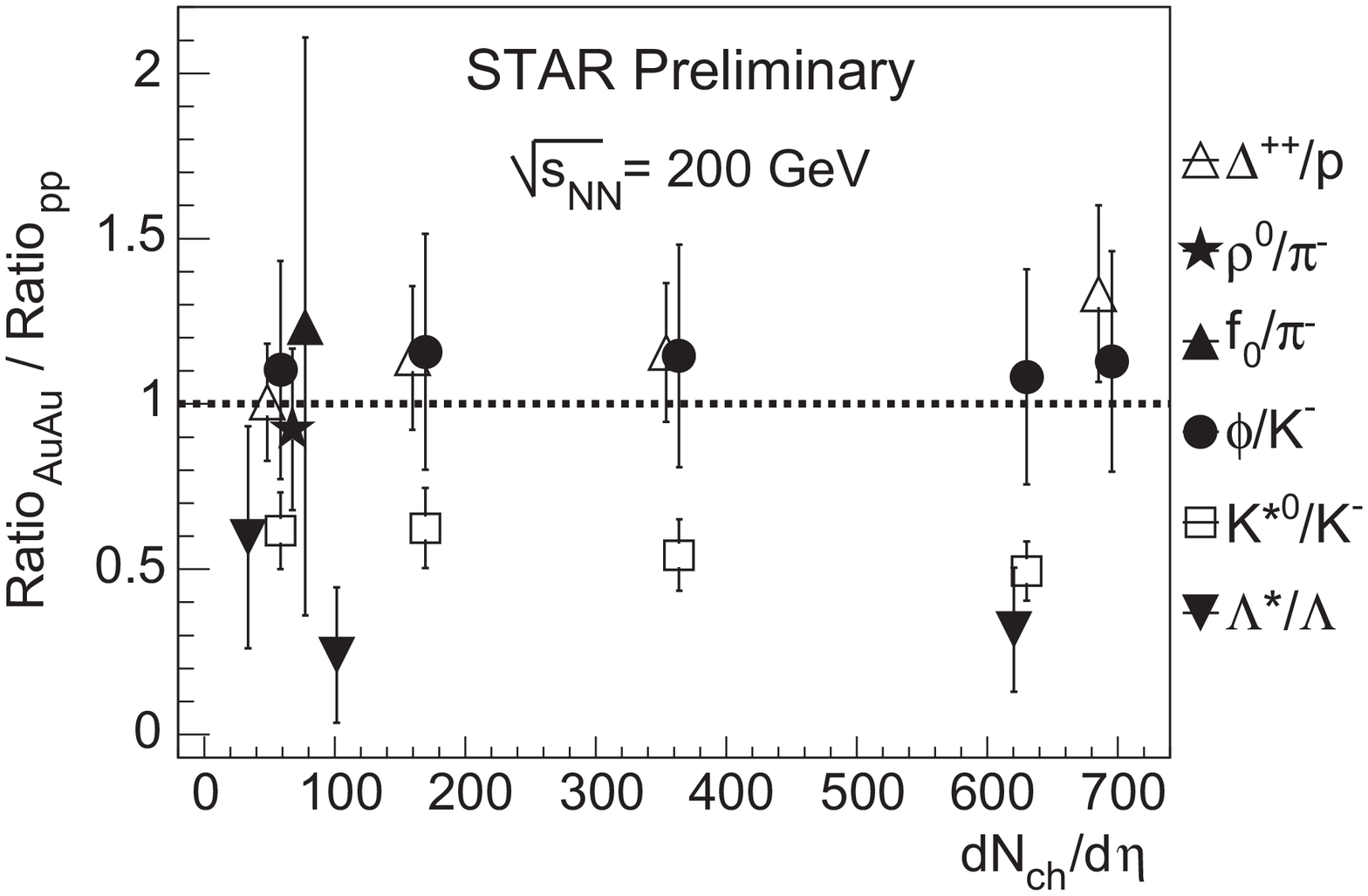}
\end{minipage}
\hspace{\fill}
\begin{minipage}[t]{85mm}
\includegraphics[height=14pc,width=17pc]{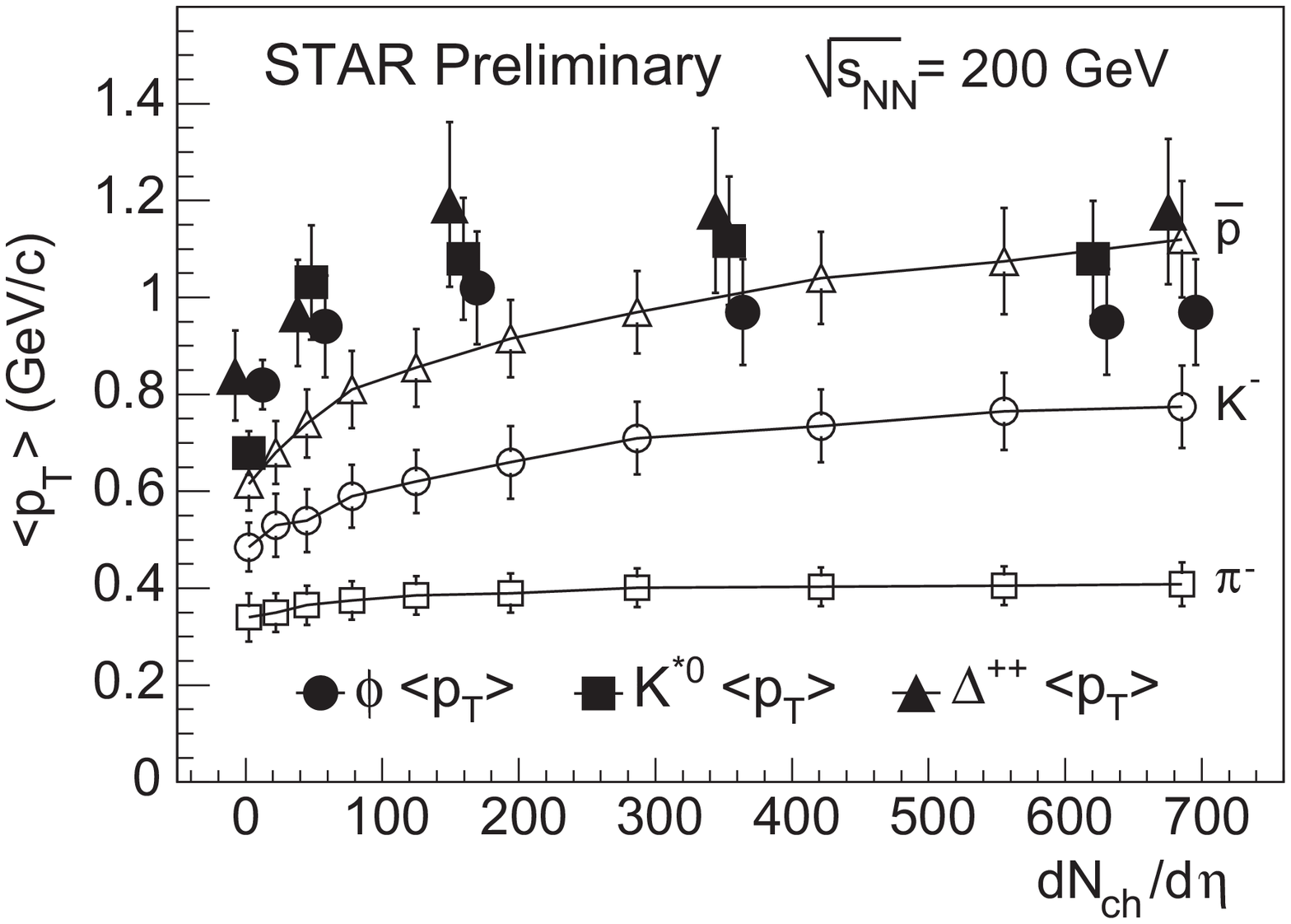}
\end{minipage}
\caption{Left panel: The $\rho^0/\pi^-$, $K^{*0}/K^-$,
$f_{0}/\pi^-$, $\phi/K^-$, $\Delta^{++}/p$, and
$\Lambda^*/\Lambda$ ratios as a function of $dN_{ch}/d\eta$ at
$\sqrt{s_{_{NN}}}\!=\!$ 200 GeV measured by STAR. All ratios have
been normalized to the corresponding ratio measured in minimum
bias $p+p$ collisions at the same c.m. system energy and indicated
by the dashed line. Right panel: The $K^{*0}$, $\phi$, and
$\Delta^{++}$ $\langle p_T\rangle$ as a function of
$dN_{ch}/d\eta$ compared to that of $\pi^-$,$K^-$, and $\bar{p}$.
The errors shown are the quadratic sum of the statistical and
systematic errors.} \label{fig:Ratio3}
\end{figure}

Figure~\ref{fig:Ratio3} depicts the $\rho^0/\pi^-$, $K^{*0}/K^-$,
$f_{0}/\pi^-$, $\phi/K^-$, $\Delta^{++}/p$, and
$\Lambda^*/\Lambda$ ratios as a function of $dN_{ch}/d\eta$ at
$\sqrt{s_{_{NN}}}\!=\!$ 200 GeV measured by STAR. All ratios have
been normalized to the corresponding ratio measured in minimum
bias $p+p$ collisions at the same $\sqrt{s_{_{NN}}}$ and indicated
by the dashed line in Fig.~\ref{fig:Ratio3}. As mentioned
previously and shown in Fig.~\ref{fig:Ratio2}, the $K^{*0}/K^-$
ratio for the top 10$\%$ of the inelastic hadronic Au+Au
cross-section is lower than the minimum bias $p+p$ measurement at
the same c.m. system energy by a factor of 2. In addition,
statistical model calculations \cite{6,7,46} are considerably
larger than the measurement presented in Fig.~\ref{fig:Ratio3}.
The $K^{*0}$ regeneration depends on $\sigma_{K\pi}$ while the
rescattering of the daughter particles depends on
$\sigma_{\pi\pi}$ and $\sigma_{\pi p}$, which are considerably
larger (factor $\sim$5) than $\sigma_{K\pi}$ \cite{13}. The lower
$K^{*0}/K^-$ ratio measured may be due to the rescattering of the
$K^{*0}$ decay products. The $\rho^0/\pi^-$ and $f_{0}/\pi^-$
ratios from minimum bias $p+p$ and peripheral Au+Au interactions
at the same c.m. system energy are comparable. Statistical model
calculations \cite{6,7,46} for Au+Au collisions underpredict
considerably the $\rho^0/\pi^-$ and $f_{0}/\pi^-$ ratios presented
in Fig.~\ref{fig:Ratio3}. The larger $\rho^0/\pi^-$ ratio measured
may be due to the interplay between the rescattering of the
$\rho^0$ decay products and $\rho^0$ regeneration. The
rescattering of the $\phi$ decay products and the $\phi$
regeneration should be negligible due to the $\phi$ longer
lifetime ($\sim$44 fm/$c$) and the small $\sigma_{KK}$,
respectively. As a result, statistical model calculations
\cite{7,46} reproduce the $\phi/K^-$ ratio measurement depicted in
Fig.~\ref{fig:Ratio3}. The centrality dependence of the $\phi/K^-$
ratio disfavors the kaon coalescence production mechanism for
$\phi$ mesons. Rescattering models \cite{47} predict an increase
in the $\phi/K^-$ ratio as a function of centrality; however, the
measurement does not support such behavior, as can be observed in
Fig.~\ref{fig:Ratio3}. The $\Delta^{++}/p$ ratio depicted in
Fig.~\ref{fig:Ratio3} has the opposite behavior than the
$K^{*0}/K^-$ ratio. While the $K^{*0}/K^-$ ratio decreases from
minimum bias $p+p$ to Au+Au interactions, the $\Delta^{++}/p$
ratio remains constant, or even slightly increase from minimum
bias $p+p$ to central Au+Au interactions. In the case of the
$\Delta^{++}$, the regeneration probability should be higher than
the rescattering of the decay products since $\sigma_{\pi p} >
\sigma_{\pi\pi}$ \cite{13,15}. Statistical model calculations
\cite{7,46} for Au+Au collisions underpredicts the $\Delta^{++}/p$
ratio presented in Fig.~\ref{fig:Ratio3}. The larger
$\Delta^{++}/p$ ratio measured may be due to the interplay between
the rescattering of the $\Delta^{++}$ decay products and
$\Delta^{++}$ regeneration.

The centrality dependence of the resonance ratios depicted in
Fig.~\ref{fig:Ratio3} suggests that the $\phi$ regeneration and
the rescattering of the $\phi$ decay products are negligible, and
the $\Delta^{++}$ regeneration is slightly larger than the
rescattering of the $\Delta^{++}$ decay products. In addition, the
results shown in Fig.~\ref{fig:Ratio3} also suggest that the
rescattering of the $K^{*0}$ decay products is dominant over the
$K^{*0}$ regeneration and therefore the reaction channel $K^*
\leftrightarrow K\pi$ is not in balance. As a result, the
$K^{*0}/K^-$ ratio can be used to estimate the time between
chemical and kinetic freeze-outs. Assuming that the minimum bias
$p+p$ measurement corresponds to the production at chemical
freeze-out and using the most central measurement of the
$K^{*0}/K^-$ ratio in Au+Au collisions for the production at
kinetic freeze-out, the time between chemical and kinetic
freeze-outs is short ($\sim$3 fm/$c$).

The $K^{*0}$, $\phi$, and $\Delta^{++}$ average transverse
momentum ($\langle p_T\rangle$) as a function of $dN_{ch}/d\eta$
are compared to that of $\pi^-$, $K^-$, and $\bar{p}$ in
Fig.~\ref{fig:Ratio3}. The $K^{*0}$, $\phi$, and $\Delta^{++}$
$\langle p_T\rangle$ do not present a significant centrality
dependence. This is contrary to the general behavior of $\pi^-$,
$K^-$, and $\bar{p}$ $\langle p_T\rangle$ that increase as a
function of $dN_{ch}/d\eta$, as expected if the transverse radial
flow of these particles increases.

\section{Conclusions}
Recent results on resonance production in A+A and $p+p$ collisions
at SPS and RHIC energies were presented. The measured $\rho^0$,
$K^{*0}$, and $\Delta^{++}$ masses are lower than previous
measurements reported in \cite{15} while the $\Delta^{++}$ width
is larger than the average reported in \cite{15}. Dynamical
interactions with the surrounding matter, interference between
various scattering channels, phase space distortions, and
Bose-Einstein correlations are possible explanations for the
apparent modification of resonance properties. The centrality
dependence of resonance ratios may be interpreted in the context
of hadronic cross sections. Using the $K^{*0}/K^-$ ratio, the time
between chemical and kinetic freeze-outs was estimated to be short
($\sim$3 fm/$c$). Further studies of resonances can provide
important information on the dynamics of relativistic collisions
and help in understanding the properties of nuclear matter under
extreme conditions.

\section*{Acknowledgements}
The author would like to thank M. Bleicher, P. Braun-Munzinger, W.
Broniowski, G.E. Brown, W. Florkowski, P. Kolb, G.D. Lafferty, F.
Laue, R. Longacre, S. Pratt, R. Rapp, E. Shuryak, T. Ullrich, Z.
Xu, and H. Zhang for valuable discussions.


\section*{References}
\begin{thebibliography}{9}
\bibitem{1} Rapp R and Wambach J 2000 {\it Adv. Nucl. Phys.} {\bf 25} 1.
\bibitem{2} Barz H W {\it et al.} 1991 {\it Phys. Lett.} B {\bf 265} 219.
\bibitem{3} P. Braun-Munzinger, Private communication.
\bibitem{4} Shuryak E V and Brown G E 2003 {\it Nucl. Phys.} A {\bf 717} 322.
\bibitem{5} Kolb P F and Prakash M 2003 {\it Preprint} nucl-th/0301007.
\bibitem{6} Rapp R 2003 {\it Nucl. Phys.} A {\bf 725} 254.
\bibitem{7} Broniowski W {\it et al.} 2003 {\it Preprint} nucl-th/0306034.
\bibitem{8} Bleicher M and St\"ocker H 2004 {\it J. Phys.} G {\bf 30} S111.
\bibitem{9} Pratt S and Bauer W 2003 {\it Preprint} nucl-th/0308087.
\bibitem{10} Granet P {\it et al.} 1978 {\it Nucl. Phys.} B {\bf 140} 389.
\bibitem{11} Ayala A {\it et al.} 2004 {\it Preprint} hep-ph/0403220.
\bibitem{12} Longacre R S 2003 {\it Preprint} nucl-th/0303068.
\bibitem{13} Protopopescu S D {\it et al.} 1973 Phys. Rev. D {\bf 7} 1279.
\bibitem{14} Matison M J {\it et al.} 1974 Phys. Rev. D {\bf 9} 1872.
\bibitem{15} Hagiwara K {\it et al.} 2002 {\it Phys. Rev.} D {\bf 66} 010001.
\bibitem{16} Adams J {\it et al.} 2003 {\it Preprint} nucl-ex/0305034.
\bibitem{17} Zhang H 2004 {\it J. Phys.} G {\bf 30} S577; {\it Preprint} nucl-ex/0403010.
\bibitem{18} Fachini P 2004 {\it J. Phys.} G {\bf 30} S565.
\bibitem{19} Ma J 2004 {\it J. Phys.} G {\bf 30} S543.
\bibitem{20} Markert C 2004 These proceedings.
\bibitem{21} Salur S 2004 {\it Preprint} nucl-ex/0403009.
\bibitem{22} Seto R 2004 These proceedings.
\bibitem{23} Gazdzicki M 2004 These proceedings.
\bibitem{24} Acton P D {\it et al.} 1992 {\it Z. Phys.} C {\bf 56} 521;
Lafferty G D 1993 {\it Z. Phys.} C {\bf 60} 659.
\bibitem{25} Ackerstaff K {\it et al.} 1998 {\it Eur. Phys. J.} C {\bf 5} 411.
\bibitem{26} Buskulic D {\it et al.} 1996 {\it Z. Phys.} C {\bf 69}
379.
\bibitem{27} Aguilar-Benitez M {\it et al.} 1991 {\it Z. Phys.} C {\bf 50} 405.
\bibitem{28} Albrecht H {\it et al.} 1994 {\it Z. Phys.} C {\bf 61} 1.
\bibitem{29} Derrick M {\it et al.} 1985 {\it Phys. Lett.} B {\bf 158} 519.
\bibitem{30} Pei Y J {\it et al.} 1996 {\it Z. Phys.} C {\bf 72} 39.
\bibitem{31} Blobel V {\it et al.} 1974 {\it Phys. Lett.} B {\bf 48} 73.
\bibitem{32} Singer R {\it et al.} 1976 {\it Phys. Lett.} B {\bf 60} 385.
\bibitem{33} Drijard D {\it et al.} 1981 {\it Z. Phys.} C {\bf 9} 293.
\bibitem{34} Chliapnikov P V {\it et al.} 1980  {\it Nucl. Phys.} B {\bf 176} 303.
\bibitem{35} Winkelmann F C {\it et al.} 1975 {\it Phys. Lett.} B {\bf 56} 101.
\bibitem{36} Chun S and Buchanan C 1998 {\it Phys. Rep.} {\bf 292} 239.
\bibitem{37} Barish K N {\it et al.} 2003 {\it Phys. Rev.} C {\bf 67} 014902.
\bibitem{38} Aggarwal M M {\it et al.} 2000 {\it Phys. Lett.} B {\bf 477} 37.
\bibitem{39} Back B B {\it et al.} 2003 {\it Preprint} nucl-ex/0304017.
\bibitem{40} Afanasiev S V {\it et al.} 2000 {\it Phys. Lett.} B {\bf 491} 59.
\bibitem{41} Abe K {\it et al.} 1999 {\it Phys. Rev.} D {\bf 59} 052001.
\bibitem{42} Akesson T {\it et al.} 1982  {\it Nucl. Phys.} B {\bf 203} 27.
\bibitem{43} Canter J {\it et al.} 1979 {\it Phys. Rev.} D {\bf 20} 1029.
\bibitem{44} Adler C {\it et al.} 2002 {\it Phys. Rev.} C {\bf 65}
041901(R).
\bibitem{45} Adler C {\it et al.} 2002 {\it Phys. Rev.} C {\bf 66}
061901(R).
\bibitem{46} Braun-Munzinger P {\it et al.} 2001 {\it Phys. Lett.} B {\bf 518} 41;
Stachel J and Magestro D, Private communication.
\bibitem{47} Bleicher M {\it et al.} H 1999 {\it J. Phys.} G {\bf 25} 1854;
\end{thebibliography}
\end{document}